\def   \ni {\noindent}

\def   \ssk {\vskip  5truept}

\def   \bsk {\vskip 15truept}

\def   \newline {\hfil\break}

\documentstyle[epsfig]{article}
\begin{document}

\hsize 6truein
\vsize 9truein
\topmargin -0.5in
\oddsidemargin 0.25in
\evensidemargin 0.25in
\font\abstract=cmr8
\font\keywords=cmr8
\font\caption=cmr8
\font\references=cmr8
\font\text=cmr10
\font\affiliation=cmssi10
\font\author=cmss10
\font\mc=cmss8
\font\title=cmssbx10 scaled\magstep2
\font\alcit=cmti7 scaled\magstephalf
\font\alcin=cmr6
\font\ita=cmti8
\font\mma=cmr8
\def\ref{\par\noindent\hangindent 15pt}
\null
%\vskip 3.0truecm
%\baselineskip = 12pt

% ------ beginning of font "title" ------

\title{\ni Supernova Science with an Advanced Compton Telescope}
\bsk \bsk
\author{\ni P.A.~Milne $^{1}$,
R.A. ~Kroeger $^{2}$, L.-S. ~The $^{3}$}
\bsk
\affiliation{1} {NRC/NRL Resident Research Associate,
Naval Research Lab, Code 7650,
Washington DC 20375}

\affiliation{2} {Naval Research Lab, Code 7650,
Washington DC 20375}

\affiliation{3} {Clemson University, Clemson, SC 29634}

\bsk
\baselineskip = 12pt

% beginning of font "abstract and keywords"
\abstract{ABSTRACT \ni
Gamma-ray line emission is a direct probe of the nucleosynthesis 
that occurs in Type Ia supernovae. In this work we describe the 
wealth of information obtainable from observations of this emission. 
Advanced Compton telescope designs are being studied by the 
Naval Research Laboratory, with the goal being the construction of 
a telescope which would be capable of 
detecting SNe Ia to distances in excess of 100 Mpc. We describe 
the instrument capabilities and the design issues that are being addressed.
We assume a SN Ia rate and quantify the frequency at which an 
advanced Compton telescope could detect, discriminate between, and 
diagnose Type Ia supernovae. From these estimates, 
we argue that an advanced Compton telescope
would be a powerful astrophysical tool. 

}
\bsk
\baselineskip = 12pt

\bsk
\baselineskip = 12pt

\text{\ni 1. INTRODUCTION}
\ssk
\ni

A supernova (SN) is the brilliant death of a star. Due to 
the thermonuclear explosion of a CO white dwarf (SN Ia) or to the 
core-collapse of a massive star (SN II/Ib/Ic), supernovae (SNe) alter the 
composition of the progenitor object, impulsively synthesizing new 
isotopes. It was recognized in the 1960s that the decay of radionuclei 
was required to explain the light curves of 
supernovae (Colgate et al. 1966, Truran et al. 1967, Bodansky et al. 1968). 
 The principal radionuclei studied 
have been $^{56}$Ni and its daughter product $^{56}$Co, but $^{57}$Co, 
$^{44}$Ti, $^{22}$Na 
and $^{60}$Co have also been suggested to contribute to 
the optical light (Woosley, Pinto \& Hartmann 1989). 
%Fransson, Houck \& Kozma 1996)
The study of SN nucleosynthesis has branched into 
three principle categories; explosive nucleosynthesis, radiation 
transport, and galactic chemical evolution. The first category of study
concentrates upon applying nuclear physics and magneto-hydrodynamics to 
simulate the evolution of the progenitor object(s) up to and through explosive 
nucleosynthesis. The results from these studies are; a) the determination 
of whether (or for what range of parameters) a given explosion scenario 
successfully yields a SN explosion, and b) the composition and kinematic 
structure of the ejecta of successful explosions, especially the yields 
of radionuclei. The second category of study accepts the outputs of the 
first group (SN models) and seeks to derive further constraints upon SNe 
through comparisons with observations. The key physics involved with these 
investigations are the transport of the decay products of the various 
radionuclei ($\gamma$/x-ray photons and positrons) and the subsequent diffusion 
and emission of UV/OPT/IR photons. The third category of study attempts to 
reproduce the local, galactic and extra-galactic abundances of all isotopes. 
The information 
critical to the third study is the exact nucleosynthesis contribution from 
SNe as a function of time and galaxy location. 

The study of gamma-ray line emission is an excellent diagnostic of SNe 
which can contribute to all three investigations. 
In the following three sections, we discuss the science derived from gamma-ray 
line observations. In the first section, 
we discuss the physics of nuclear decays and the escape of gamma-ray photons. 
In the second section, we describe the specifications of an advanced Compton telescope (ACT), 
particularly the version being investigated at the Naval Research Lab (NRL). 
In the third section, we discuss the science that 
could be performed with an ACT. We conclude with an 
assessment of what we consider to be the priorities of SN science with an ACT.

\begin{figure}
\begin{center}
\centerline{
\hspace{3.3cm}
\epsfig{file=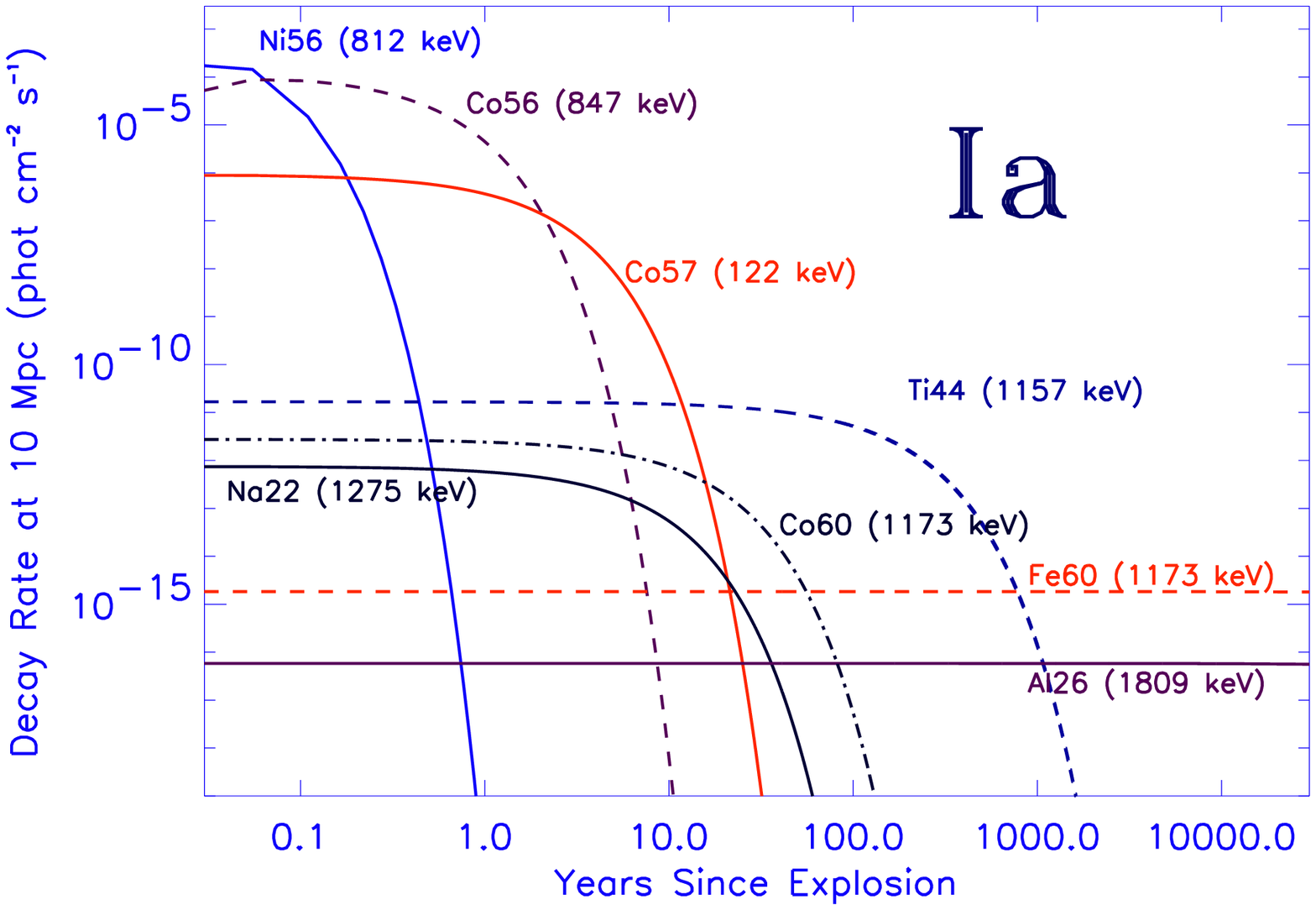, width=2.9in}
\hspace{0.2cm}
\epsfig{file=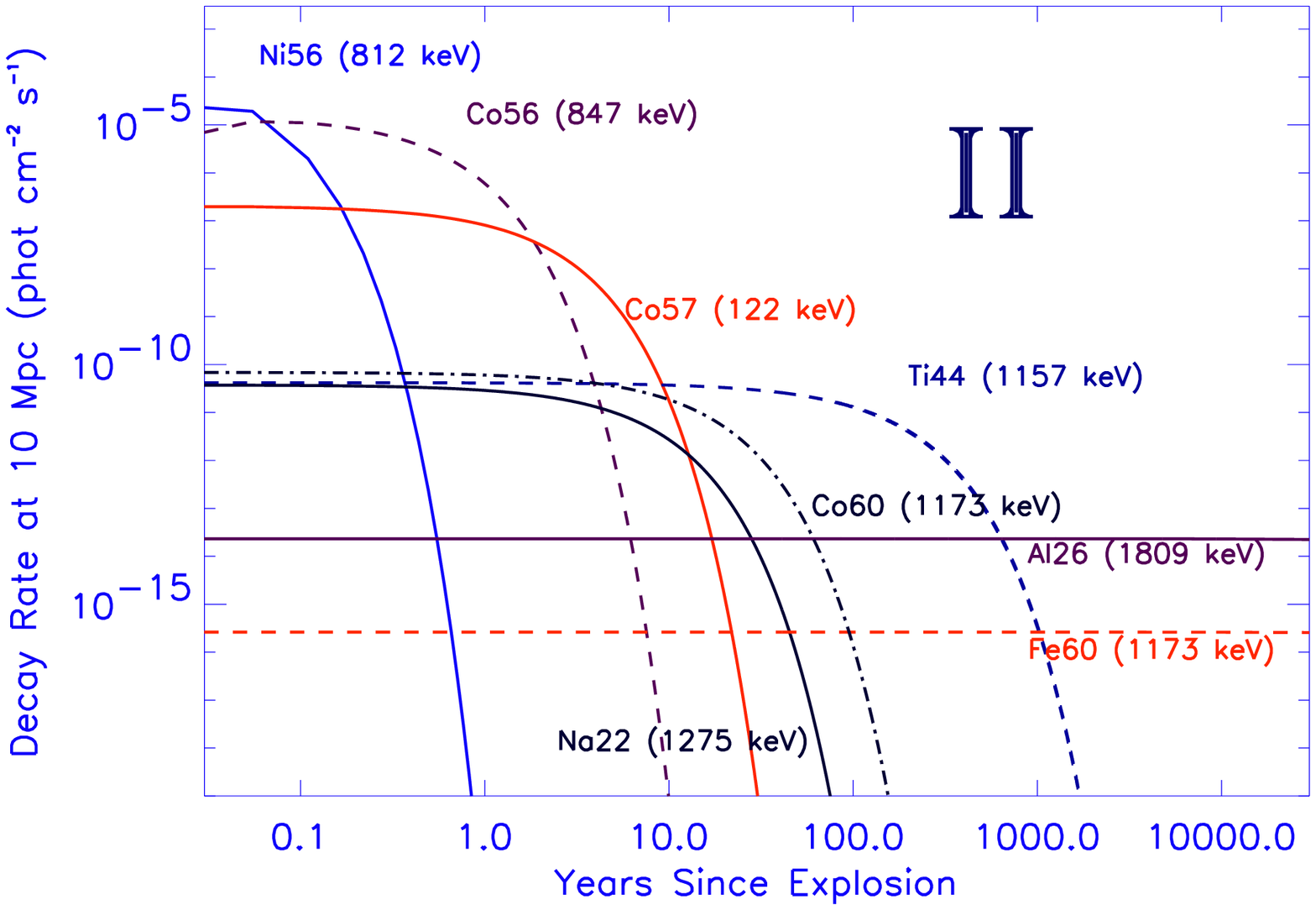, width=2.9in}
}
\vspace{-.0cm}
\caption{{\bf Figure 1.} The decay rates of the SN Ia model, W7, 
and the SN II model, W10HMM. Both SNe are at a distance of 10 Mpc.}
The large scale hides the difference of a factor of eight in the 
$^{56}$Ni production between the two SN types.
\end{center}
\end{figure}

\section{Gamma-Ray Line Emission Physics}

 Radionuclei produced in SNe decay to stable nuclei on various time-scales, 
generating gamma- and x-ray photons, electrons 
 and positrons. These decay products either 
deposit their energy in the ejecta (at early times helping to lift the 
ejecta, at later times driving the UV/OPT/IR light curves) or they escape, 
leading to potentially detectable $\gamma$/x-ray emission. Shown in Figure 1 
are the decay curves for the 
principal radionuclides of two SN models.\footnote{This would be the flux 
observed if 100\% the photons escape ({\it i.e.} zero opacity).}\footnote{ 
The decay of $^{56}$Ni $\rightarrow$ $^{56}$Co produces 750 \& 1562 keV lines 
in addition to the 812 keV line. The decay of $^{56}$Co $\rightarrow$ 
$^{56}$Fe produces a 1238 keV line in addition to the 847 keV line. The 
branching ratios are 0.49 (750), 0.86 (812), 0.14 (1562), 1.00 (847), 
0.68 (1238) (Nadyozhin 1994).}  
The left panel shows the decay rates of the SN Ia model  
W7 (Nomoto et al. 1984). The right panel shows the decay 
rates for the SN II model  W10HMM (Pinto \& Woosley 1988). 
Evident in this figure is
%Pinto & Woosley, Nature, 333, 534 (1988)
the cascade from the early-time dominance of short-lived radioactivities 
to the later dominance of long-lived radioactivities. Assuming that a large 
fraction of these photons escape, short-lived radioactivities give rise to 
intense, but brief emission, while long-lived radioactivities give rise to 
faint, but persistent emission. For the purposes of this work, the 
gamma emission is categorized as prompt, supernova remnant (SNR), or diffuse. This 
categorization is based upon both the physics of the gamma-ray emission 
(the opacity and the angular size of the emitting region) and upon the 
instruments employed to detect the emission. 

Initially, the SN density is so large that all $\gamma$/x -ray photons are 
scattered and no high-energy emission emerges. As the SN expands, the 
ejecta thins and the $\gamma$/x -ray photons begin to escape. Thus, the 
{\em prompt} $\gamma$/x -ray line flux from a SN depends upon the overlying 
mass and the ejecta kinematics. This makes the evolution of the fluxes of
the various gamma-ray lines a probe of the SN ejecta. The optical 
absorption and emission lines from SNe also probe these quantities, but not 
as early, nor as directly as do the gamma-ray lines. The {\em prompt} epoch 
is characterized by the lowering of the gamma-ray line opacity to 
negligible values. The early onset of gamma-ray escape would make 
prompt emission detectable to large distances. Although plausible SNe Ia and 
SNe II/Ib/Ic explosion scenarios exhibit a range of characteristics, 
the timescale of prompt emission is on the order of a year.

\begin{figure}
\begin{center}
\centerline{\epsfig{file=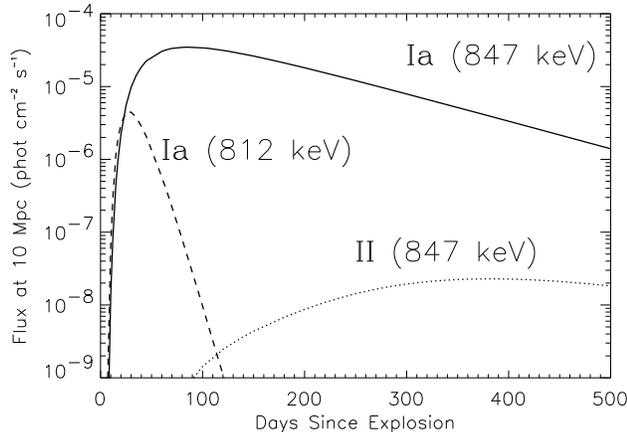, width=3.0in}}
\vspace{-.0cm}
\caption{{\bf Figure 2.} 
The 812 \& 847 keV line fluxes for the same Type Ia and 
Type II models shown in Figure 1.
}
\end{center}
\end{figure}

Two different approaches have been employed to detect SN gamma rays; wide-FoV 
instruments, and pointed, target of opportunity (ToO) instruments. 
For a wide-FoV detector, the temporal and 
sky-coverage are large enough that the desired detection rate is achieved 
through low sensitivity and long-duration observations. This search 
discovers SNe independent of optical SN searches and is capable of detecting 
any escaping emission from the moment of explosion. This approach was 
employed by the SMM/GRS which detected 847/1238 keV emission from 
SN 1987A (Matz et al. 1988).\footnote{The diagnostic utility of 
prompt emission was exemplified 
by the SMM/GRS studies of SN 1987A. The line appeared earlier than expected 
which suggested that $^{56}$Ni may have been mixed outward in the ejecta. 
The total $^{56}$Ni agreed with values suggested by light-curve 
studies.} Narrow FoV detectors rely upon triggers (optical 
detections) to observe ToO SNe. 
These instruments are engineered to have maximal 
sensitivity as a trade-off for the narrower FoV. The critical element in this 
approach is the time delay between the optical SN detection (and SN 
type identification) and the re-orientation of the gamma-ray instrument to 
observe the SN. 

As is evident in Figure 1, the dominant prompt emissions 
are the $^{56}$Ni $\rightarrow$ $^{56}$Co decays and the subsequent $^{56}$Co 
$\rightarrow$ $^{56}$Fe decays. Type II SNe produce less $^{56}$Ni than type 
Ia SNe (compare the peaks of the 812 \& 847 keV lines in Figure 1) and have 
the $^{56}$Ni buried under many solar masses of ejecta. For both of these 
reasons, prompt emission is far fainter in SNe II than SNe Ia. Figure 2 
shows the time evolution of the line fluxes of the 812 \& 847 keV lines for
the same models shown in Figure 1. The SN Ia lines are much more 
intense than the SN II lines. Although the SN II rate is $\sim$ 5 times 
larger than the SN Ia rate, the far larger flux for SNe Ia make them the 
dominant SN type for studies of {\em prompt} gamma-ray emission. For SNe Ia, 
the 812 keV line peaks sometime within the first 30 days after the 
explosion. ToO instruments would need to initiate observations no later than 
13 days before the peak optical luminosity to fully study that feature. 
Current optical SN searches do not regularly detect SNe that 
early.\footnote{The difficulties of relying upon optical triggers was 
demonstrated by the CGRO/COMPTEL \& OSSE observations of SN 1998bu. 
The SN was discovered nine days before peak, but CGRO observations 
began eight days later. The SN was observed for 88$^{d}$, but was not 
detected in gamma-ray line emission (Georgii et al. 1999).} 
SN searches would need to be dramatically improved to facilitate that 
requirement. A wide FoV instrument does not require an optical 
or any other external trigger.

The scientific insight that can be attained by studying the {\em prompt} emission 
from SNe Ia is the topic of this work, but longer-lived emissions 
are worthy of mention. The fluxes shown in Figure 2 suggest that 
{\em prompt} emission from SNe Ia can be detected to distances on the 
order of 100 Mpc by a detector capable of 10$^{-7}$ photon cm$^{-2}$ s$^{-1}$ sensitivitity. 
Barring a fortuitous galactic SN during the mission, 
prompt emission studies will investigate extra-galactic SNe.

Radionuclei with longer life-times emit at lower flux levels, but the 
emission from the nuclei may persist for centuries. The lower flux 
levels dictate that young SNRs can only be detected out to the Local Group. 
Older SNRs must be galactic, but the emission can be detected on decadal- 
millenial time-scales. {\em SNR} studies thus 
concentrate upon $^{57}$Co(122 keV), $^{22}$Na(1275 keV), 
$^{60}$Co(1173,1332 keV), and $^{44}$Ti(68,78,1157 keV). The 
$^{57}$Co(122 keV) line was detected from SN 1987A, a young SNR in the LMC  
by the CGRO/OSSE instrument (Kurfess et al. 1992). 
The $^{44}$Ti(1157 keV) line was detected from the galactic 
SNRs, Cas A (Iyudin et al. 1994) and RX J0852-4642 (Iyudin et al. 1998), 
by the CGRO/COMPTEL instrument.\footnote{Recent analyses have lowered the significance 
of the 1157 keV line detection from RX J0852-4642 to the 2$\sigma$ -4$\sigma$ level 
(Sch\H{o}nfelder 1999).} The SN 1987A 
detection suggests that the $^{57}$Co decay rate was too low to explain 
the 1600 day optical luminosity. The Cas A detection suggests more 
$^{44}$Ti production than was expected from the apparently low-luminosity 
SN observed by Flamsteed. The RX J0852-4642 detection motivated an 
archival search of ROSAT x-ray data to discover a ``new" young SNR 
(Aschenbach et al. 1998). 
A next-generation gamma-ray telescope will contribute to 
the study of SNRs by detecting gamma-ray emission from many more SNRs 
(both historical and undiscovered) and 
by improving the flux estimates of previously detected SNRs. In addition, 
the improved angular resolution will allow studies of the spatial 
extent of the emission for bright, distributed SNRs. 

Very long-lived radioactivities decay at relatively low rates. For 
the 
nearest SNRs, $^{26}$Al and $^{60}$Fe may be detectable similar to the 
$^{57}$Co,$^{22}$Na,$^{60}$Co and $^{44}$Ti emission. For these SNRs 
the 1809/1173 ratio is a discriminant between SN types. 
The 10$^{6}$ year life-times of these radionuclei dictate that the majority 
of detectable SNe will be too old to be resolved as individual SNRs. 
The study of this {\em diffuse} emission will investigate the 
cumulative contributions of 10,000 or more SNe. Maps of this emission are 
then the integrated million-year history of all SNe in the Galaxy (Diehl 
et al. 1995, Oberlack et al. 1996, Knodlseder et al. 1999).
The 511 keV line emission from the annihilation of positrons produced
in the decays of $^{56}$Co, $^{44}$Ti and $^{26}$Al will also produce a
diffuse emission (see Chan and Lingenfelter 1993 for a review of 
potential positron sources). 
Studies of diffuse emission contribute important 
information for models of galactic chemical evolution.

\section{Estimate Parameters}

\subsection{Specifications of an ACT}

NASA's Gamma-Ray Working Group (GRAPWG) has identified (June 1999) 
the study of 
nuclear astrophysics and sites of gamma-ray line emission as its 
highest priority science topic, and the development of an Advanced 
Compton Telescope (ACT) as its highest priority major mission. 
This instrument will follow three other instruments optimized to 
perform gamma-ray SN studies. The first generation 
Compton telescope was the COMPTEL instrument, which operated 
on-board the Compton Gamma-Ray Observatory (CGRO) from 1991-2000. COMPTEL
used a liquid scintillator to produce a single Compton scatter 
and a NaI scintillator to capture the scattered energy. That 
combination achieved an instrumental angular resolution on the order of 
2-4 degrees rms, and an energy resolution of 5-8 \% FWHM. COMPTEL had a 
FoV of almost 60$^{\circ}$, and observed SNe as targets of opportunity 
(ToO). COMPTEL may have 
marginally detected the 847 keV and 1238 keV $^{56}$Co lines 
from the SN 1991T (Morris et al. 1997). The OSSE instrument, also on-board 
CGRO, used four NaI(Tl)-CsI(Na) scintillators to observe SNe as a ToO detector. 
OSSE had a 9\% energy resolution and used a tungsten collimator to achieve a 
3.8$^{\circ}$x11.4$^{\circ}$ FWHM FoV. OSSE observations of SN 1991T and 
1998bu did not detect either of the $^{56}$Co decay lines. 
The SPI instrument on-board the INTEGRAL satellite will employ a 
coded-aperture germanium detector to observe gamma-rays in the 20 keV 
-2 MeV energy range. SPI is expected to achieve an angular resolution 
of 2-3 degrees FWHM, and an energy resolution of 2 keV (@ 1 MeV). The 
SPI instrument will have a narrow FoV (16$^{\circ}$), 
that will be used as a ToO instrument for prompt SN science.
The IBIS instrument, also on-board the INTEGRAL satellite, will also be 
capable of detecting broad-line gamma-ray emission with senstivities almost 
equivalent to SPI.

\begin{figure}
\begin{center}
\centerline{\epsfig{file=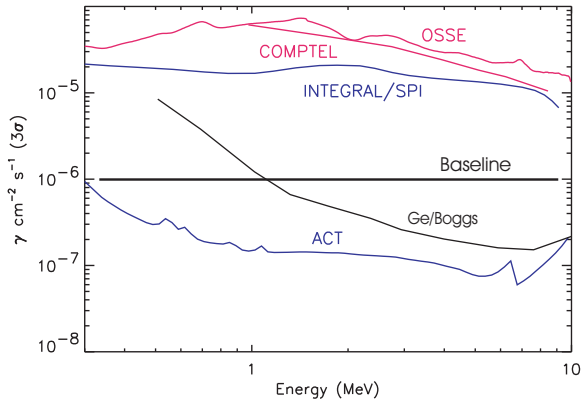, width=3.0in}}
\vspace{-.0cm}
\caption{{\bf Figure 3.}
 The sensitivity curves of the ACT and four other gamma-ray 
detectors. The curves are 3$\sigma$ for 10$^{6}$ second accumulation.
}
\end{center}
\end{figure}

The GRAPWG has outlined a baseline ACT with a goal of achieving a 
point source localization accuracy of 
of $\sim$ 0.1$^{\circ}$, an energy resolution of $\leq$ 3 keV 
(@2 MeV), a FoV of 60$^{\circ}$ and a broad-line sensitivity of 
1 x 10$^{-6}$ phot cm$^{-2}$ s$^{-1}$ (10$^{6}$s, 3$\sigma$). 
The Naval Research Laboratory (NRL) is investigating both germanium 
 and silicon Compton telescope designs, 
with the intention of exceeding the baseline 
specifications in both sensitivity and FoV (Kurfess et al. 1999). 
The advances that will make
this improvement possible are: 1) large volume detector arrays which will 
increase the effective area, 2) excellent spatial and energy 
resolution from the use of position-sensitive solid-state detectors, and 
3) employing two Compton scatters and a third interaction 
 to determine the incoming energy and 
angle rather than a single scatter and total energy absorption (as 
employed with COMPTEL). The current NRL/ACT design will 
increase the FoV to 120$^{\circ}$, and improve 
the broad-line sensitivity to 3 x 10$^{-7}$ phot cm$^{-2}$ s$^{-1}$  
(10$^{6}$s, 3$\sigma$). The point source localization 
will be poorer than the 
baseline, increasing to 0.5 degrees for the weakest sources, but 
improves for strong sources. The energy resolution is expected to 
be $\sim$ 20 keV. Boggs \& Jean (2000) performed a simulation to estimate 
the sensitivity of a germanium Compton telescope. 
The telescope sensitivity was based on a background estimate for 
high-Earth orbit, and the efficiency derived from a simulation. Good 
events are those that undergo 3 or more interactions and are totally 
absorbed. 
 Shown in Figure 3 are the broad-line 
sensitivity curves for the instruments described 
above.\footnote{The Boggs and Jean
Ge simulation was scaled to low-Earth orbit assuming a lowering of the
background by a factor of ten, and a 20 keV FWHM line width.} 
The sensitivity degrades significantly below 500 keV, making the 478 and 
511 keV lines the lowest energy lines anticipated to be detectable. 

\begin{figure}
\begin{center}
\centerline{\epsfig{file=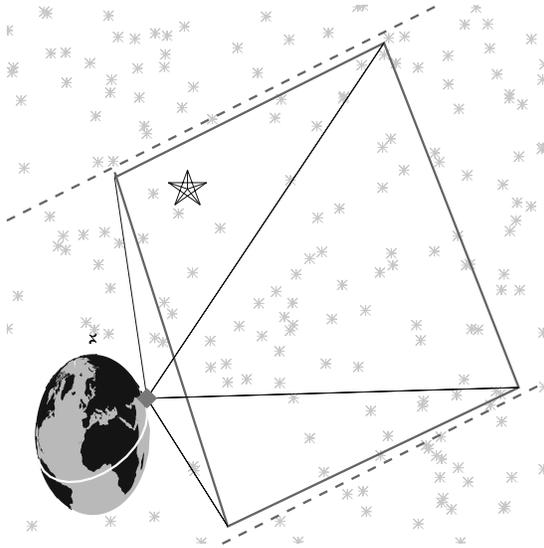, width=3.0in}}
\vspace{-.0cm}
\caption{{\bf Figure 4.} 
A schematic representation of the orbit of an ACT.
}
\end{center}
\end{figure}

If the ACT is placed in an equatorial, low-Earth orbit and 
scans along the celestial equator, as shown in Figure 4, it will observe 
roughly 87\% of the sky every orbit, with a source within the FoV 1/3 
of the time. By contrast, a 60$^{\circ}$ FoV instrument in the same orbit would
observe 51\% of the sky with one-sixth accumulation efficiency. Included in 
the FoV will be the Virgo Cluster (a likely region for SNe detections) and
the Galactic Center (for studies of galactic gamma-ray sources).
This strategy has two advantages. For prompt emissions, the telescope will 
be able to observe the SN {\em before} optical
 discovery, thus collecting photons 
on the rising and fading sides of the peaks.
For SNR emissions, the telescope will revisit the remnant every orbit over
the entire mission. This will lead to a composite observation with 
exposure far in excess of the 10$^{6}$s typical for a ToO observation.

Ideally, a wide-FoV optical instrument would accompany the Compton 
telescope, observing the same fields. The optical monitor would insure 
that all SNe would be observed every orbit from the time of explosion. 
Larger, ground-based telescopes would respond to gamma and optical 
triggers, providing follow-up spectra, light curves and precise 
positions. This would insure that all SNe detected in gamma-rays would be 
well-studied in the standard wavelengths.

\subsection{SN Ia Rate}

SN rates are on the order of a few SNe per century per galaxy, 
meaning that many 
galaxies must be surveyed to produce a useable number of detections. 
SN searches have systematically observed galaxies and attempted to 
quantify the selection effects of their searches to derive the 
actual SN rate. SN rates are quoted in ``SNu", or the number of 
SNe per century per unit of blue luminosity (10$^{10}$ L$_{\odot}^{B}$). 
SN rates depend upon the value of the Hubble constant, in this work we 
assume H$_{o}$=68 km s$^{-1}$ Mpc$^{-1}$. Hamuy \& Pinto (1999) estimated 
the rate to be {\it R}=0.23$^{+0.33}_{-0.14}$ SNu h$_{68}^{2}$. 
Cappellaro et al. (1997) 
estimated the rate to be {\it R}=0.18$\pm$0.06 SNu h$_{68}^{2}$. Hardin et al. 
(2000) estimated the rate to be {\it R}=0.20$^{+0.16}_{-0.09}$ 
SNu h$_{68}^{2}$.
Averaging these estimates, we assume the rate to be {\it R}=0.2 SNu.
To convert this value to a rate per volume, the luminosity density must be 
known. Hamuy \& Pinto (1999) quoted Marzke, claiming the luminosity 
density to be L$_{B}$ = (1.12 $\pm$ 0.29) x 10$^{-2}$ h$_{68}$ 
(10$^{10}$ L$_{\odot}^{B}$ Mpc$^{-1}$).\footnote{Lin et al. 1996 claim a 
lower value, L$_{B}$ = (0.95 $\pm$ 0.10) x 10$^{-2}$ h$_{68}$
(10$^{10}$ L$_{\odot}^{B}$ Mpc$^{-1}$).} Combining these estimates 
suggests there are {\em 100 SNe Ia per year within 100 Mpc}. 

The 812 \& 847 keV line fluxes shown in the next section suggest that SNe Ia 
will be detectable out to 100 -200 Mpc. The previous studies cited all 
sampled galaxies within roughly the same volume.\footnote{The 
Cappellaro et al. (1997) search had a mean recession velocity of $z \sim$ 
0.01 (64 Mpc). The 
Hardin et al. (2000) search was the deepest, and it 
operated out to $z \sim$ 0.1 (440 Mpc).} 
We will assume that the SNe Ia density is uniform throughout the 
entire volume for the following estimates.

SNe Ia are not a homogeneous class. Individual SNe have been both brighter 
and fainter than ``normal" SNe. Li et al. (1999) suggests that only 60\% 
of SNe Ia are normally-luminous (N), with 20\% super-luminous (SP) and 
20\% sub-luminous (sb). We will use that 3:1:1 ratio in this work. 

\begin{figure}
\begin{center}
\centerline{\epsfig{file=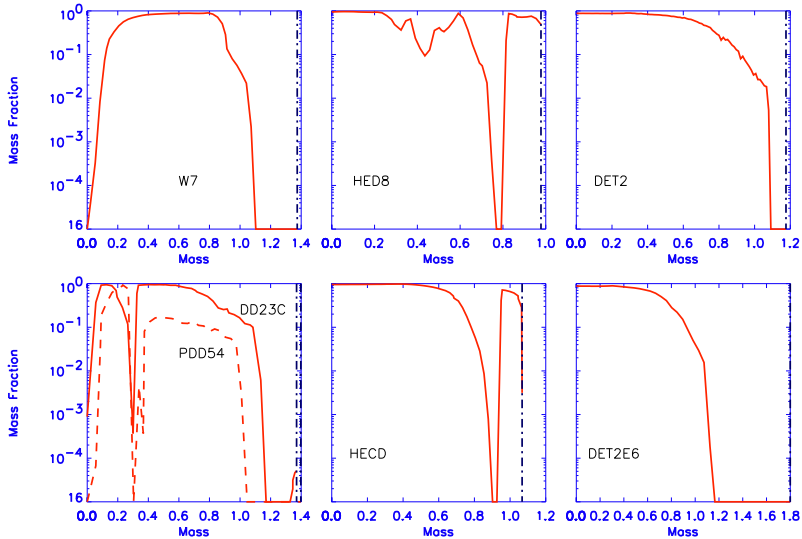, width=4.0in}}
\vspace{-.0cm}
\caption{{\bf Figure 5.} The radial distribution of $^{56}$Ni 
for seven SN Ia models. The masses are in units of solar masses. 
The dashed lines denote the outer surface of the ejecta.
}
\end{center}
\end{figure}

\subsection{SN Ia Models}

Considerable variations exist within the collection of SN Ia models. 
These variations are driven both by the uncertainty as to the correct 
explosion scenario(s) and by the heterogeneity displayed in the spectra 
and light curves of observed SNe Ia. While strong arguments have been 
made for and against each explosion scenario, the single-degenerate, 
Chandrasekhar mass scenario (SDCM) has emerged as the most 
plausible (Wheeler 1995, Livio 2000, Pinto \& Eastman 2000b) .
%\cite{livi99,pint00,whee99} 
We have included four SDCM models in 
our simulations; (1) the normally-luminous deflagration, W7, 
(2) the normally-luminous delayed detonation, DD23C (H\H{o}flich et al. 1998), 
(3) the sub-luminous pulsed-delayed detonation, PDD54 (H\H{o}flich, Khokhlov 
\& Wheeler 1995), and (4) the super-luminous delayed detonation, 
W7DT (Yamaoka et al. 1992). The peak 
spectra of sub-Chandrasekhar mass models (SC) have been suggested to be too 
blue to explain SNe Ia (Nugent et al. 1995, H\H{o}flich \& Khokhlov
1996), but the range of progenitor masses can naturally 
account for the peak-width/luminosity relation (Pinto \& Eastman 2000a). 
We have included three variants of the 
SC scenario; (1) 
the normally-luminous SC model, HED8 (0.96 M$_{\cdot}$ (M$_{ej}$), 0.51 
M$_{\cdot}$($^{56}$Ni)), (2) the sub-luminous SC model, HED6 
(0.77 M$_{\cdot}$ (M$_{ej}$), 0.26 M$_{\cdot}$($^{56}$Ni),
 H\H{o}flich \& Khokhlov 1996), (3) the super-luminous SC model, 
HECD (1.07 M$_{\cdot}$ (M$_{ej}$), 0.72 M$_{\cdot}$($^{56}$Ni) 
Kumagai \& Nomoto 1997). 
A double degenerate scenario involving the merger of two CO white dwarfs 
has also been suggested to explain SNe Ia. 
Disagreements as to whether this ``double degenerate" (DD) scenario leads
 to a SN Ia or an accretion-induced collapse (AIC) are 
on-going. An equally serious objection has been the failure to detect 
enough progenitor systems, although recent studies hint that enough 
DD SNe Ia could occur. We have included three variants of the DD 
scenario; DET2, DET2ENV2, 
and DET2ENV6.\footnote{H\H{o}flich et al. 1996 has shown that 
DD models do not follow the peak-width/luminosity relation.} The $^{56}$Ni 
distributions for 7 of the 10 models are shown in Figure 5. The DD models 
bury the $^{56}$Ni under 0.6 -1.2 M$_{\cdot}$ of ejecta, whereas SC models 
feature large amounts of $^{56}$Ni near the surface. A small variation 
is shown within the SDCM group, where DD23C (and W7DT) have $^{56}$Ni 
nearer the surface than does W7.  

The gamma-ray line photons from the $^{56}$Ni \& $^{56}$Co decays are 
either Compton scattered to lower energy or escape the ejecta. We 
simulate the scattering adopting the prescription of Podznyakov, Sobol \& 
Sunyaev (1983). A detailed description of the Monte Carlo algorithm and 
its application in calculating the spectra and bolometric light curves 
of SN 1987A and Type Ia SNe have been presented by The, Burrows \& Bussard 
(1990) and Burrows \& The (1990). 
Shown in Figure 6 are the 812 \& 847 keV line fluxes for the ten models, 
as would be observed from 10 Mpc. The 812 keV line peaks early and is a 
probe of the mass overlying the outermost $^{56}$Ni-rich ejecta. The 
847 keV line peaks later (at which time the ejecta for most models has 
become optically thin) and probes the total $^{56}$Ni production. The solid 
lines are line fluxes calculated in this work, dashed lines and dot-dashed 
lines are results from other investigators. It is clear that our results, 
as well as Pinto \& Eastman's results, predict lower fluxes than either 
H\H{o}flich, Khokhlov \& Wheeler (1998) (hereafter HKW), or Kumagai \& 
Nomoto (1997).
Our 847 keV line fluxes converge to the HKW fluxes at late times. 
Additionally, the energy deposition rates of HKW match well our calculations. 
Current efforts are being made to understand and resolve the discrepancies. 
Note that our results suggest fewer SN detections than the HKW and Kumagai 
results would suggest. 

\begin{figure}
\begin{center}
\centerline{\epsfig{file=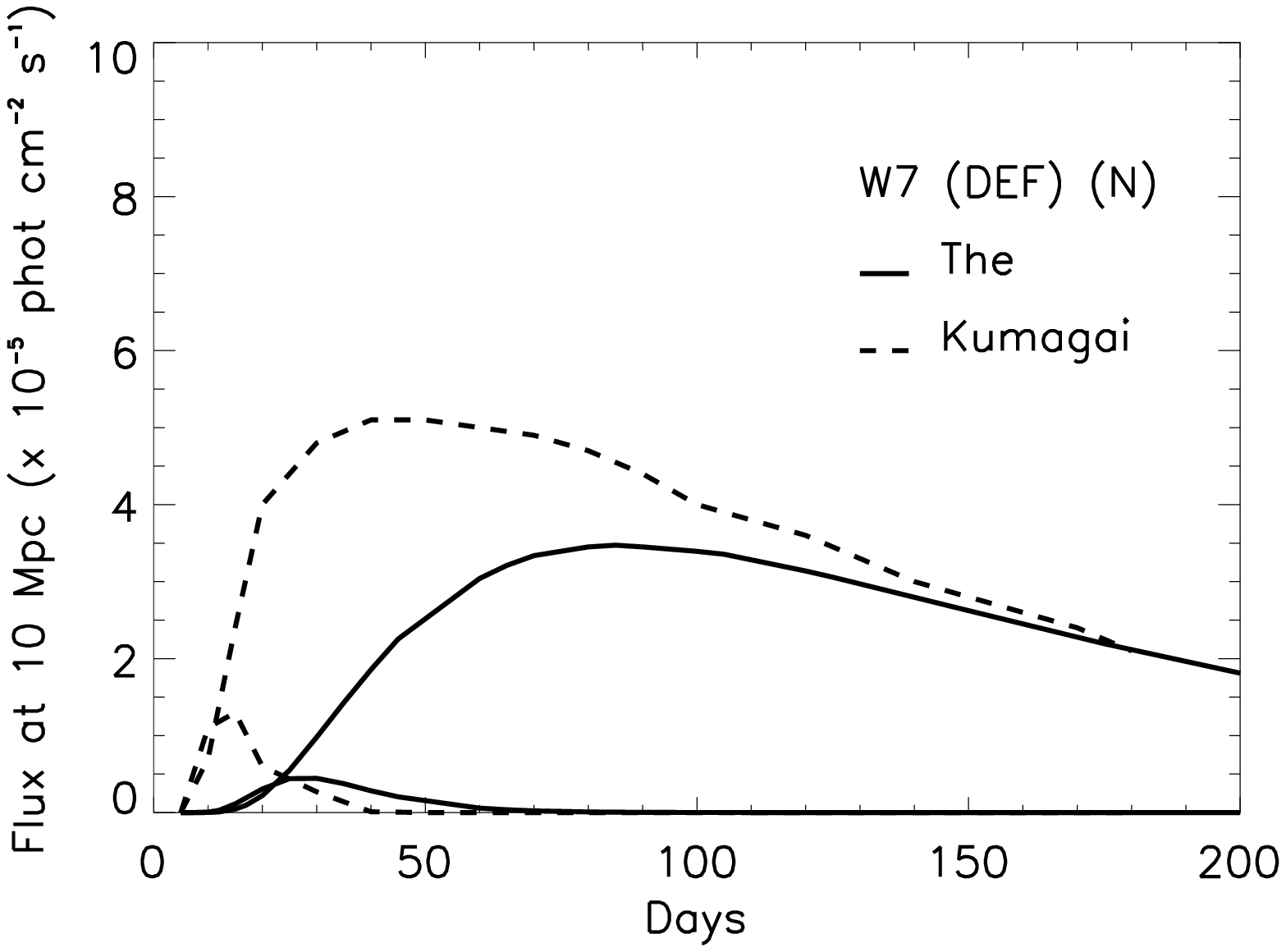, width=3.0in}
\epsfig{file=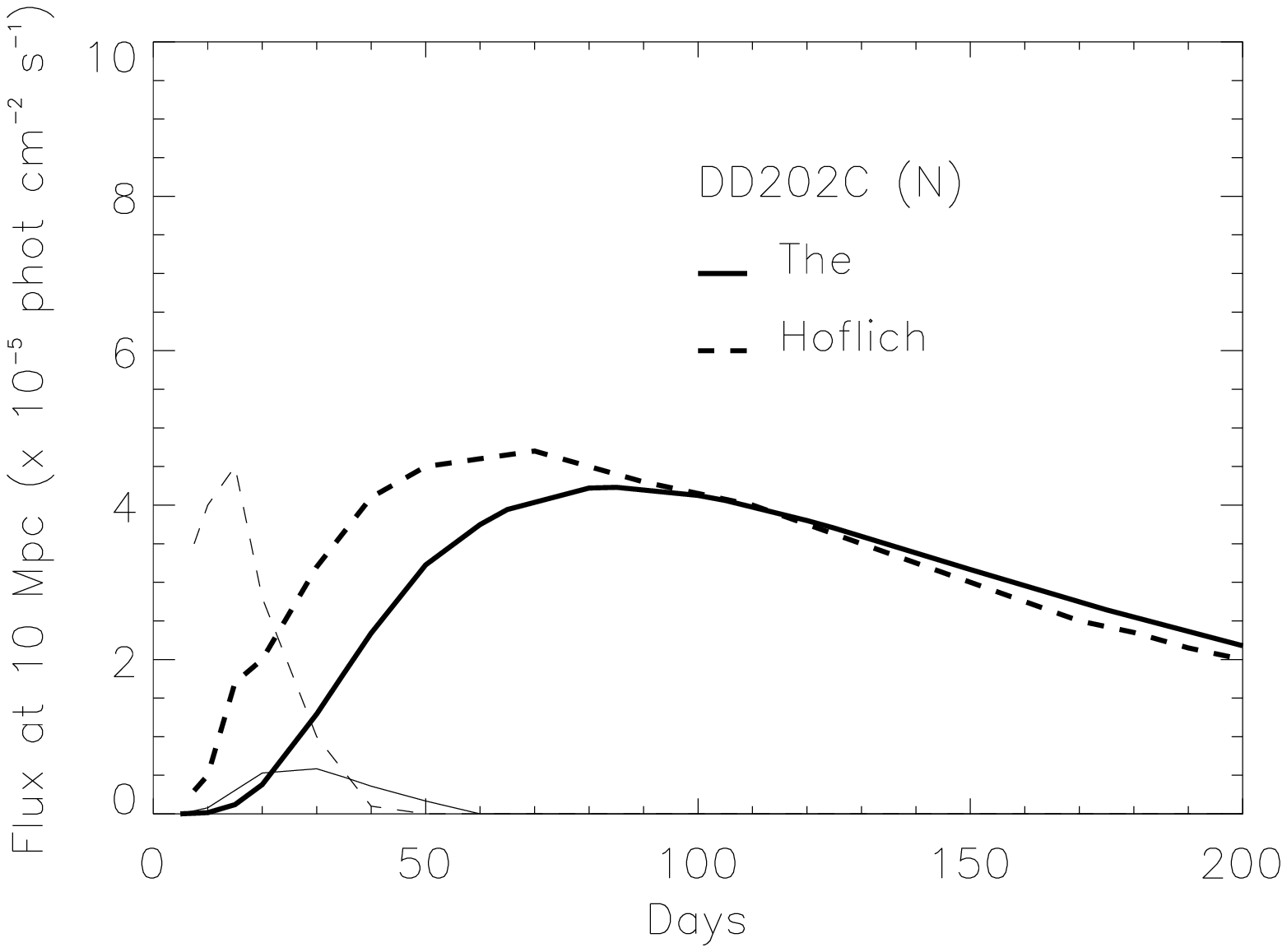, width=3.0in}
}

\centerline{
\epsfig{file=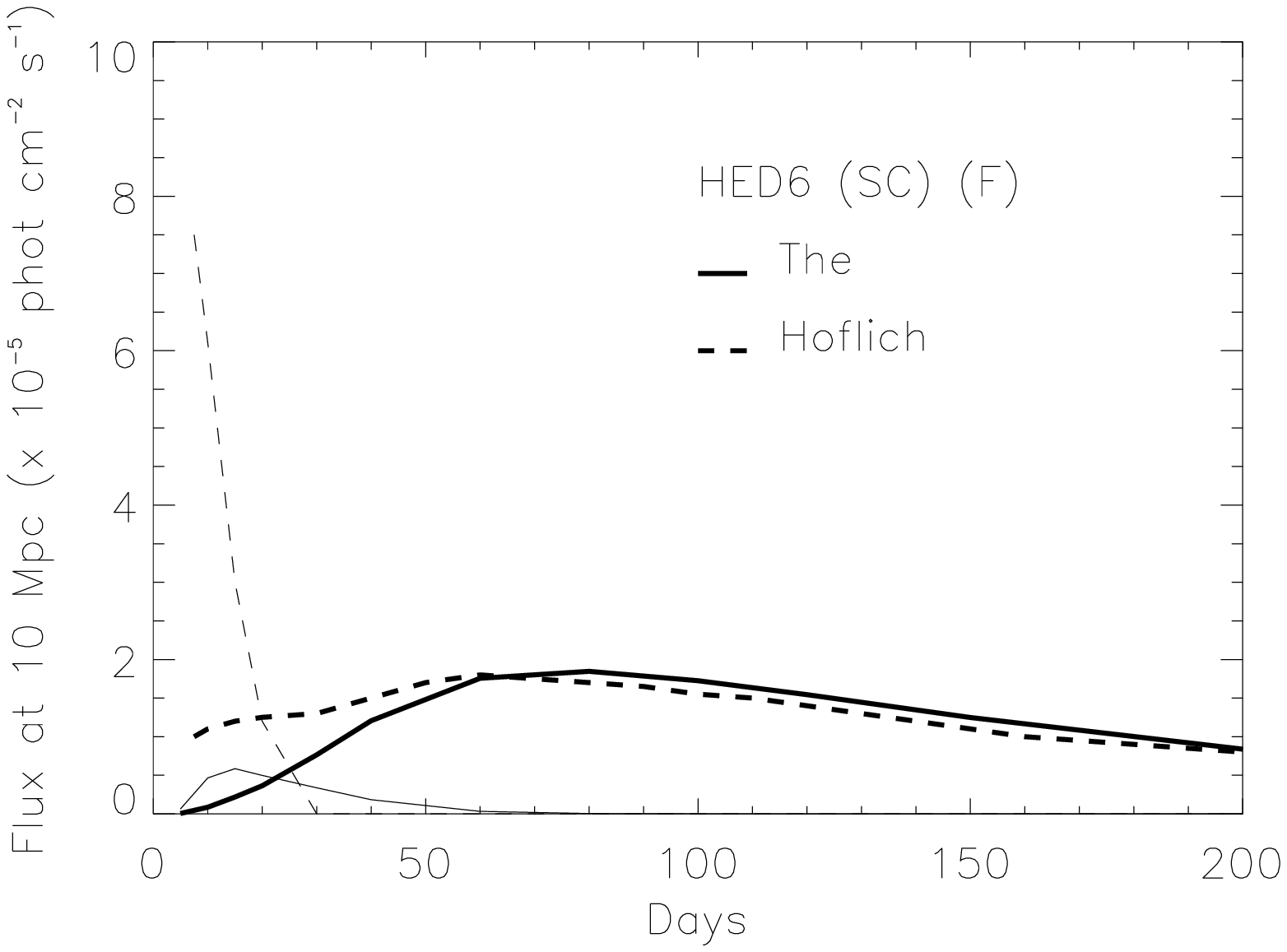, width=3.0in}
\epsfig{file=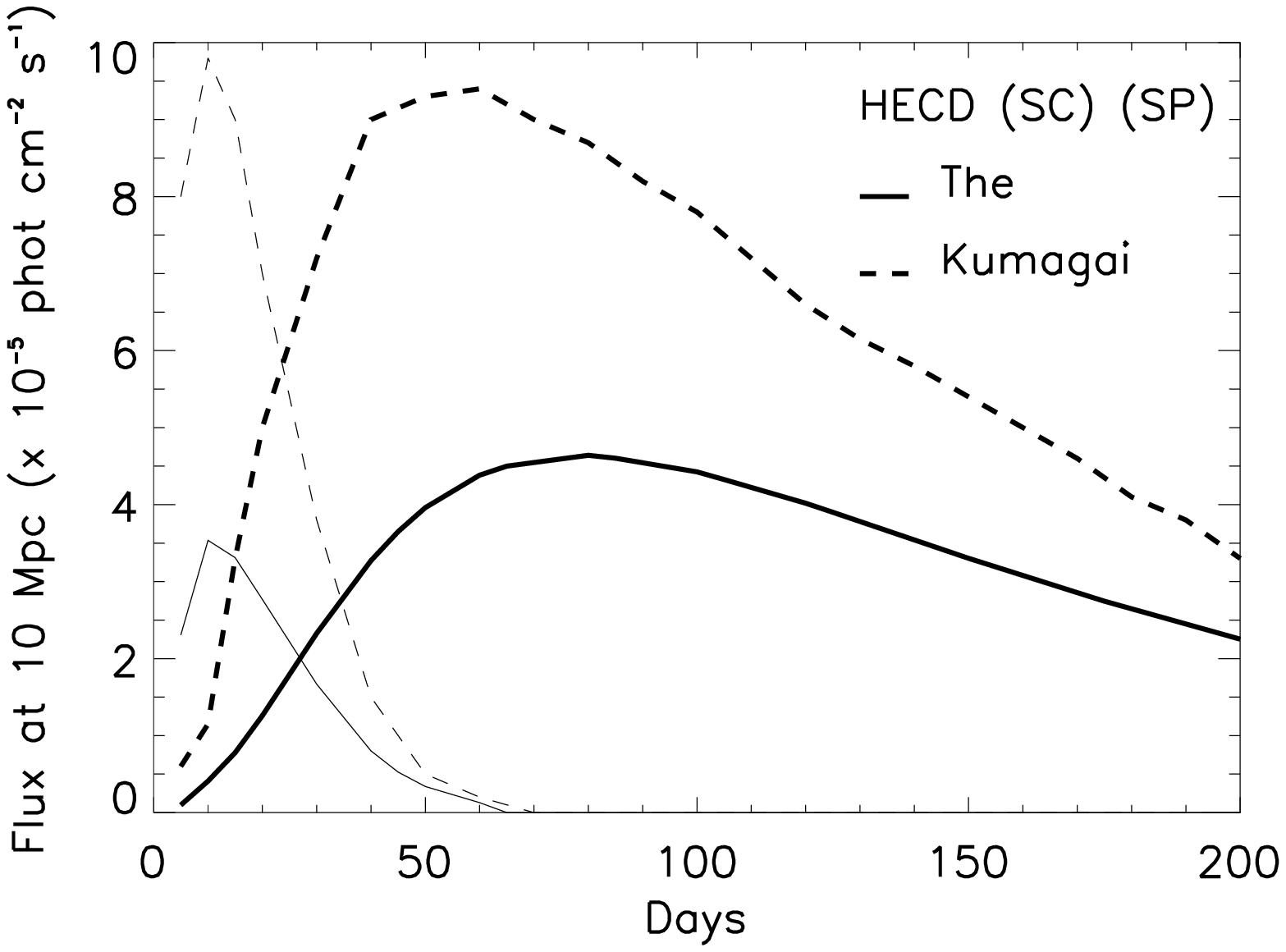, width=3.0in}
}

\centerline{
\epsfig{file=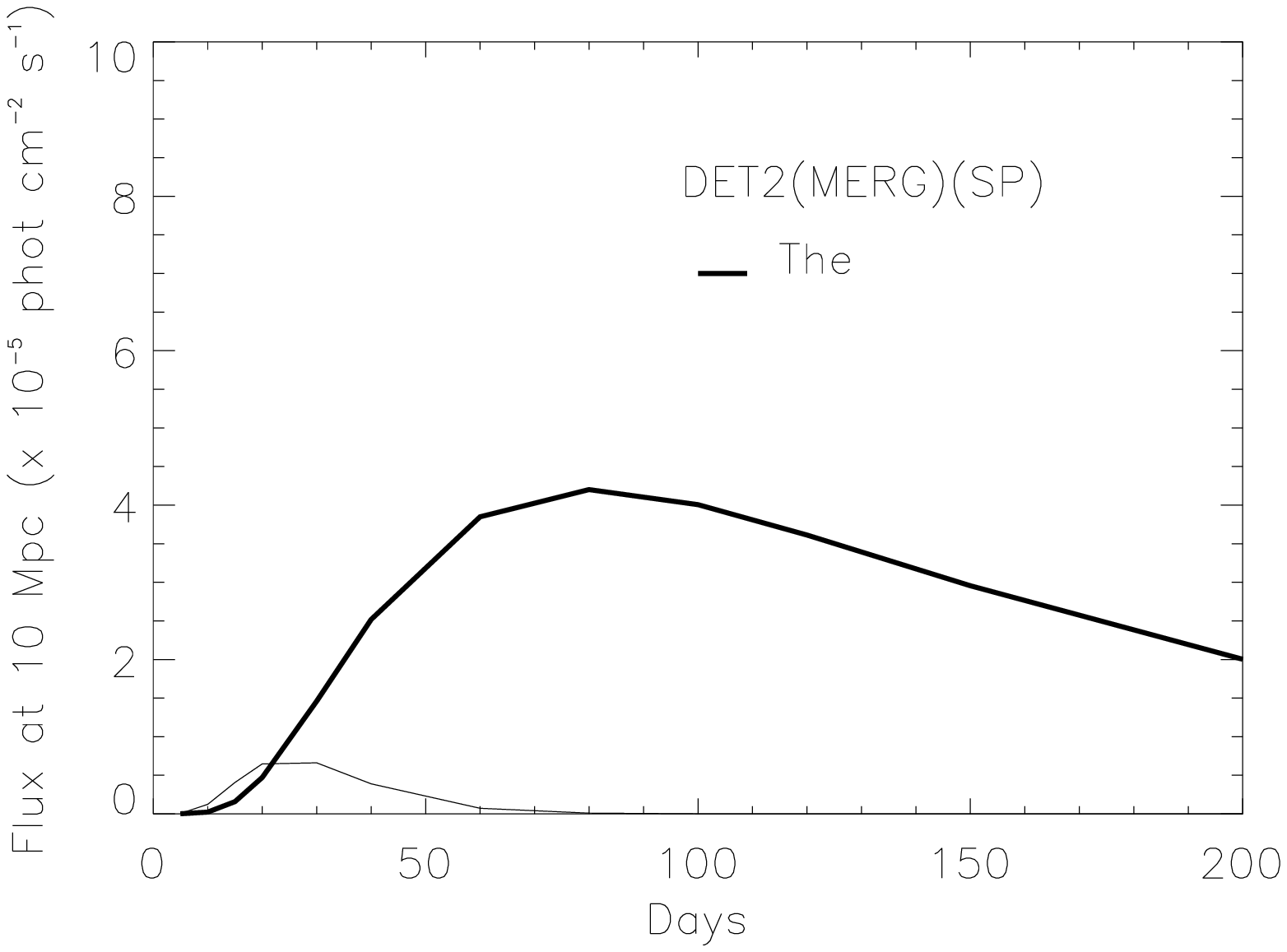, width=3.0in}
\epsfig{file=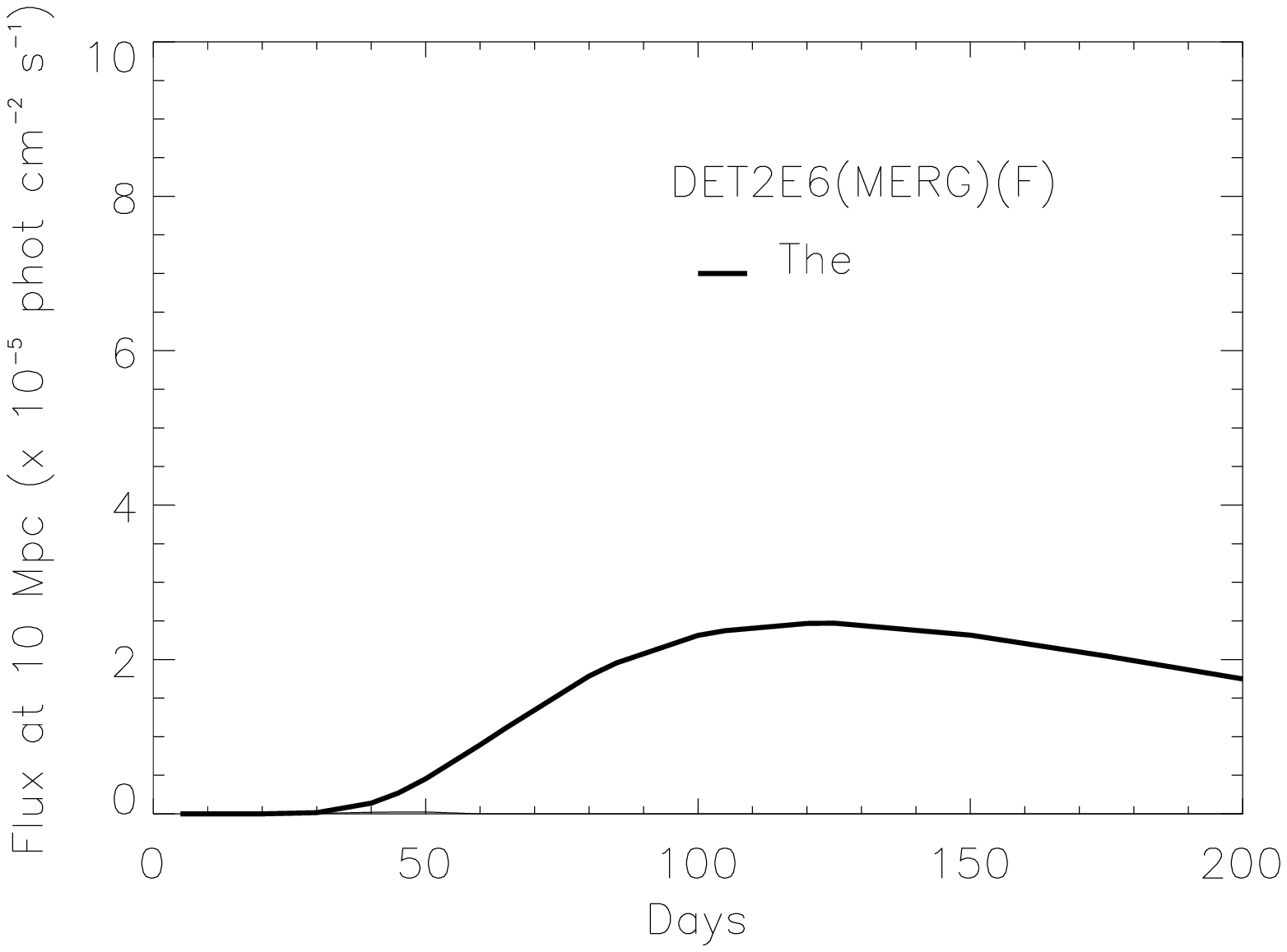, width=3.0in}
}
\vspace{-.0cm}
\caption{{\bf Figure 6.}
Gamma-ray lines fluxes for six SN Ia Models at 10 Mpc. 
The thin curves show 812 
keV line fluxes, the thick lines show 847 keV line fluxes. The solid 
lines were calculated for this work, the dashed lines are from 
HKW or Kumagai \& Nomoto 1997.}
\end{center}
\end{figure}
                  
Concentrating upon our results, there are clear differences between the 
explosion scenarios. The flux ratio of the 812 keV line 
peak to the 847 keV line peak is a 
distance-independent discriminant. DD models feature very low 
812/847 ratios, SC models feature high ratios. When combined with 
estimates of the host galaxy's distance, the 847 keV peak flux can determine 
the $^{56}$Ni production and thus discriminate between SDCM \& SC 
scenarios. In the next section, we quantify these tendencies to suggest 
the rate at which different levels of science can be achieved with an ACT.

\section{SN Ia Science with an ACT}

An ACT will be capable of performing various degrees of scientific 
investigation of prompt emission from SNe Ia, depending upon the SN distance.
For this work, we categorize the science as {\em detections, discrimination, 
and diagnostics}, and describe each category separately. 

\subsection{Detections}

Gamma-ray detection of SNe Ia will be the most basic level of science 
that will be performed with an ACT. Through SN detections we learn about the 
relative rates of SN Ia sub-classes (N,SP,sb), the SN Ia rate as a 
function of galaxy morphological class, and the projected radial 
distribution of SNe Ia. All of these studies assume that the gamma-ray 
observations are coordinated with optical observations. By detecting SNe 
both optically and in gamma-rays, we can get a measure of the selection 
biases of each SN search. Howell et al. (1999) demonstrates that optical 
SN searches (visual, photographic \& CCD) suffer from the ``Shaw effect", 
where SNe in the inner bulges of galaxies are less frequently detected. 
Additionally, the optical emission from SNe suffers extinction from 
intervening dust. Highly extincted SNe are less likely to be detected. 
Gamma-rays are not expected to suffer from these effects and will detect 
all SNe Ia in a galaxy. This information is particularly relevant to 
determining the SN contribution to galactic chemical evolution.

\begin{table}
\begin{center}
\caption{{\bf TABLE 1.} Maximum detectable distance and 
detection rates for various SN Ia models.$^{a,b}$}
\begin{tabular}{lllll}
\\
\hline
\hline
\vspace{0mm}
Model & \multicolumn{2}{c}{GRAPWG Baseline} & 
\multicolumn{2}{c}{NRL ACT Concept} \\
Name  & Distance [Mpc] & Rate [SNe/yr] & Distance [Mpc] & Rate [SNe/yr] \\ 
\hline
Normal & & & & \\
W7 & 57 & 5.8 & 123 & 97 \\
HED8 & 60 & 6.5 & 128 & 110 \\
DD23C & 56 & 5.3 & 120 & 89 \\
\hline
Sub-lum & & & & \\
W7DT & 69 & 3.4 & 148 & 57 \\
HECD & 67 & 3.1 & 145 & 53 \\
DET2 & 62 & 2.5 & 134 & 42 \\
\hline
Super-lum & & & & \\
PDD54 & 30 & 0.3 & 65 & 4.8 \\
HED6 & 41 & 0.7 & 89 & 12 \\
DET2E6 & 50 & 1.1 & 103 & 19 \\
\hline
%Type II & \multicolumn{2}{c}{750 kpc} & \multicolumn{2}{c}{1.6 Mpc} \\
Type II & 750 kpc &--- & 1.6 Mpc &--- \\
\hline
\hline
\multicolumn{5}{l}{a Detections are at the 5$\sigma$ level.} \\
\multicolumn{5}{l}{b Based on gamma-ray line fluxes shown in Figure 6.}\\
\end{tabular}
\end{center}
\end{table}

Many SNe need to be detected for the estimated SN rates and distributions 
to be significant. Shown in Table 1 are the distances and rates at which 
given SN models could be detected (5$\sigma$) via the combined observation 
of the 750, 812, 847, 1238 \& 1562 keV lines. The estimates are made for 
both the GRAPWG baseline specifications and the enhanced sensitivity
 design being 
studied at NRL. For each ACT, the $^{56}$Ni-rich super-luminous SNe Ia will 
be detected to the largest distances, but the larger SN rate of 
normally-luminous SNe Ia make them the most frequently sampled sub-class. 
The NRL-ACT will detect SNe Ia to roughly twice the distance as would the 
baseline ACT, 
and will detect $\sim$ 160 SNe Ia/yr compared to $\sim$ 10 SNe 
Ia/yr. We assert that a 5 year sampling of 800 SNe Ia will allow SN rate 
studies at a scientifically-interesting level, while 50 SNe Ia are 
inadequate.\footnote{The NRL-ACT design detects more SNe than the baseline 
ACT due to a wider FoV and a better sensitivity.}

Detecting SNe with a wide FoV instrument depends upon the {\it integrated 
flux}. The differences between the simulations shown in Figure 6 occur 
principally at early times. Using the larger fluxes suggested by 
HKW and Kumagai would not lead to an appreciable increase in the number of 
detections. 

\subsection{Discrimination}

For a subset of the detected SNe Ia, the line fluxes will be large 
enough to generate light curves. This will allow discrimination between 
the various models suggested to explain each sub-class. The estimated 
rate (per year) at which a given SN model could be distinguished from 
alternative models is shown in Table 2. This discrimination assumes that 
the explosion date is known to within $\pm$3 days from optical spectra 
obtained as part of a coordinated study. 
The NRL-ACT would be able to 
discriminate between SDCM models and SC models for 15-21 SNe Ia per year 
(or 75-100 over a 5 year mission liftime). 
The deflagration (W7) could be distinguished from the delayed detonation 
(DD23C) once per year. Lower SN rates for SP and sb SNe Ia mean that 
SDCM and SC models will be distinguished 6-9 times for each sub-class 
by the better ACT (over a 5 year mission liftime). By contrast, the 
baseline ACT will detect two SNe Ia per year with which it can 
distinguish between explosion scenarios. Again we assert that the increased 
sampling of the NRL-ACT relative to the baseline ACT will critically 
improve the useable science.

\begin{table}
\begin{center}
\caption{{\bf TABLE 2.} Rates at which pairs of SN Ia models would be 
distinguished.$^{a}$}  
\begin{tabular}{llllll}
\\
\hline
\hline
\\
\multicolumn{2}{l}{GRAPWG Baseline}  & &
\multicolumn{2}{l}{NRL ACT Concept}  & \\
\hline
& HED8 & DD23C & & HED8 & DD23C \\
W7 & 0.9 & 0.1 & W7 & 15 & 1.0 \\
HED8 & --- & 1.2 & HED8 & --- & 21 \\
\hline
     & HECD & DET2 && HECD & DET2 \\
W7DT & 0.1 & 0.3 & W7DT & 1.7 & 5.8 \\
HECD & --- & 0.5 & HECD & --- & 8.3 \\
\hline
      & HED6 & DET2E6 & & HED6 & DET2E6 \\
PDD54 & 0.1 & 0.0 & PDD54 & 1.3 & 0.8 \\
HED6 & --- & 0.4 & HED6 & --- & 7.3 \\
\hline
\hline
\multicolumn{6}{l}{a Rates are in units of [SNe/yr]. 
Detections are at the 3$\sigma$ level.} \\
\end{tabular}
\end{center}
\end{table}

This information would conclusively establish which explosion 
scenario dominates SNe Ia. For normal SNe Ia, if none of the 100 or 
so of these nearby SNe were best explained by an alternative scenario, 
than that scenario could be effectively rejected. Note that 
discrimination between SN explosion scenarios is critically dependent 
upon the gamma-ray transport simulations. It is for this application of 
an ACT that simulation descrepancies must be resolved.

\subsection{Diagnostics}

For very nearby SNe Ia, the evolution of the ejecta may be 
studied in greater detail. 
Chan \& Lingenfelter (1988), Bussard, Burrows, \& The (1989), 
Burrows \& The (1990), and  HKW demonstrated that both the line  
centroid, the line width, and the line profile of gamma-ray lines 
probe the kinematics of the ejecta and reveal  
the velocity distribution of the $^{56}$Ni in the ejecta. 
Qualitatively, the line centroid will be 
initially blue-shifted due to detection of only the gamma-ray photons 
from the approaching face on the SN. As the ejecta thins to gamma-rays, 
the line centroid moves to the energy dictated by the host galaxy's 
red-shift. Most SN Ia models have the $^{56}$Ni-rich ejecta in the 
inner regions. For these SN models, the thinning of the ejecta with time
reveals an increasing range of velocities. This produces a widening 
of the gamma-ray lines with time, with the FWHM asymptotically 
approaching the FWHM dictated by the mean $^{56}$Ni velocity. 
The only exceptions to this are the SC models dominated by 
$^{56}$Ni near the surface (HED6). Those SN models 
feature an initially broad line that subsequently narrows and then  
slowly broadens to the FWHM dictated by the mean $^{56}$Ni velocity.
Characterizing the line profile/shape in general, 
for a radioactive source distributed in a thin shell, 
the gamma line profile at maximum has a box shape, while a 
uniformly distributed source has a parabolic shape (Burrows \& The 1990). 
Quantitatively, the blue-shift decreases from $\sim$20 keV to 0 keV, 
and the line width increases from $\sim$15 keV to $\sim$30 keV. 

\begin{figure}
\begin{center}
\centerline{\epsfig{file=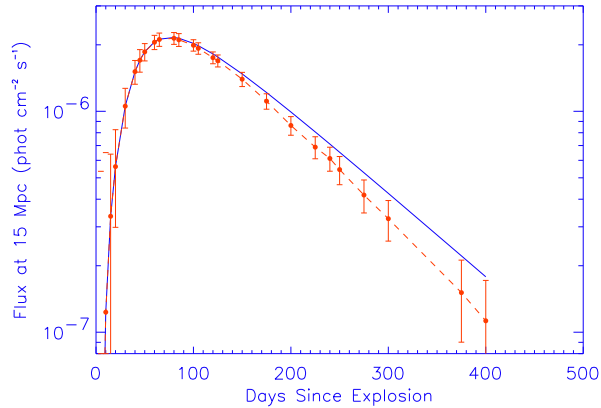, width=3.0in}}
\vspace{-.0cm}
\caption{{\bf Figure 7} The 511 keV line flux for a SN Ia at 15 
Mpc assuming two positron transport scenarios. The dashed line, 
with simulated observations, was calculated assuming positron 
escape. The solid line, which would scale with the 847 keV line, 
assumes instantaneous, in-situ positron annihilation.
}
\end{center}
\end{figure}

Whereas these effects will need to be included in the analysis to 
properly estimate line fluxes, it is uncertain whether the ACT designs 
under development will be able to discriminate between SN models 
based upon these features. The differences between the centroids are
typically less than 4 keV, the FWHM of the models vary by typically 
8 keV (the largest variation is between HED6 and PDD5, which differ 
by 20 keV FWHM at 8$^{d}$). Optimizing the ACT in light of these effects 
will be one of the challenges facing ACT designers, requiring a 
combination of high sensitivity and good energy resolution. 

Hard X-ray spectra and the K X-ray line fluxes can be used to infer  
the composition or metallicity of the ejecta 
(The, Clayton, \& Burrows 1990; The, Bridgman, \& Clayton 1994). 
X-ray band flux ratios reflect the slope of the X-ray spectrum and 
the scattering and photoelectric opacities of the ejecta. 
The time evolution of the X-ray band flux ratios and the ratios of 
the total X-ray and gamma-line luminosities can also reveal the 
extent of mixing or clumping in the ejecta.

An additional diagnostic that can be studied with an ACT is the magnetic
field of the ejecta. The decay of $^{56}$Co $\rightarrow$ $^{56}$Fe 
which produces the 847 \& 1238 keV lines also produces positrons 
in 19\% of the decays. Initially, the positrons have negligible lifetimes,  
producing two 511 keV annhilation photons that will scale with the 847 keV 
line. By 100$^{d}$, if the magnetic field is either too weak to confine 
positrons, or radially-combed permitting positrons to escape along the 
field lines, then the 511/847 line ratio will decrease as positrons escape 
the ejecta. If the field is strong and turbulent, the 511/847 ratio will 
remain constant. Milne, The \& Leising (1999), Capellaro et al. 1997, 
and Ruiz-Lapuente \& Spruit (1998) have all transported positrons through 
SN Ia models and compared model-generated optical light curves to observations 
of type Ia SNe. All three groups conclude 
that positrons may escape the ejecta, although 
disagreement exists whether this is a general result for every SN Ia. 
 Figure 7 shows the 511 keV line flux for the SN Ia model, 
HED8 at 15 Mpc. The dashed line, with simulated data, shows the line flux 
if the magnetic field is such that positrons are allowed to escape.
The solid line shows the line flux if positrons are trapped and annihilate 
in-situ (which would follow the 847 keV line). 
Assuming that the ACT sensitivity is 
6 x 10$^{-7}$ phot cm$^{-2}$ s$^{-1}$ (3$\sigma$,10$^{6}$s) 
to 511 keV broad-line emission, this could be detectable 
above the 4$\sigma$ level to 15 Mpc for the SC model, HED8, to 10 Mpc for 
the SDCM model, DD23C. 
If a very nearby SN Ia occurs, the direct determination of 
positron escape/trapping may be made.

\section{Conclusions}

The ejecta of SNe Ia is profoundly radioactive. Measuring the gamma ray 
emission from SN ejecta is a useful probe of the nucleosynthesis 
and hydrodynamics (energetic, structure, mixing/clumping) 
of the SN explosion, and thus is a discriminant between explosion 
scenarios. Although gamma-ray line emission is observable for more than a 
millennium from nearby SNRs, in this work we concentrate upon the 
science that can be realized from the study of prompt decays.  
The decays of $^{56}$Ni $\rightarrow$ 
$^{56}$Co $\rightarrow$ $^{56}$Fe produce a number of prompt gamma-ray 
lines which an ACT could detect to distances greater than 100 Mpc. The 
wide FoV of the ACT design is particularly suitable for the study of 
SNe Ia. We 
show that the current estimates of the SN Ia rate suggests that many 
SNe will be detected by such an instrument. We argue for the production 
of an ACT capable of wider FoV observations and improved sensitivity 
than the baseline design suggested by the GRAPWG. We feel that the 
rates of detection and discrimination achievable with the NRL ACT concept 
are large enough to make that instrument a landmark 
astrophysical tool. 

The time-frame of the development of an ACT is on the order of a decade. 
During that time, the field of SN Ia science will certainly advance. In 
particular, the efficiency of SN searches will appreciably improve. The 
design of the ACT must exploit the anticipated advances to best utilize 
the gamma-ray information that will be accumulated. Advances in optical 
studies will not remove the necessity of gamma-ray line observations. It 
is closer to the truth to assert that 
the combined contraints from optical and 
gamma-ray investigations will exceed the sum of the individual constraints. 
The potential of this instrument are exciting to contemplate.

\pagebreak

{\references \ni REFERENCES
\ssk

\ref Aschenbach, B. 1998, Nature, 396, 141
\ref Bodansky, D., Clayton, D.D., Fowler, W.A. 1968, ApJS, 16, 299
\ref Boggs, S.E., Jean, P.J., in Proceedings of the 4th 
INTEGRAL Workshop, in press
\ref Burrows, A. and The, L.-S. 1990, ApJ, 360, 626
\ref Capellaro, E. et al. 1997, A\&A, 322, 431
\ref Chan, K.-W., Lingenfelter, R.E. 1993, ApJ, 405, 614
\ref Colgate, S.A., White, R.H. 1966, ApJ, 143, 626
\ref Diehl, R. et al. 1995, A\&A, 298, 445
\ref Georgii, R. et al. 1999, in Proceedings of the 5th
Compton Symposium, ed. M.L. McConnell \& J.M. Ryan, 
(New York:AIP), p.49
\ref Hamuy, M., Pinto, P.A. 1999, AJ, 117, 1185
\ref Hardin, D. et al. 2000, A\&A, 
\ref H\H{o}flich, P., Khokhlov, A., Wheeler, J.C. 1995, ApJ, 444, 831
\ref H\H{o}flich, P., Khokhlov, A. 1996, ApJ, 457, 500
\ref  H\H{o}flich, P., Wheeler, J.C., Theilemann, F.-K. 1998,
ApJ, 495, 617
\ref Howell, A., Wang, L., Wheeler, J.C. 1999, astro-ph, 9908127
\ref Iyudin, A.F. et al. 1994, A\&A, 284, L1
\ref Iyudin, A.F. et al. 1998, Nature, 396, 142
\ref Knodlseder, J. et al. 1999, A\&A, 345, 813
\ref Kumagai, S., Nomoto, K. 1997, in Proceedings of the NATO ASI on 
Thermonuclear Supernovae (C486), ed. P. Ruix-lapuente, R. Canal, J. Isern, 
(Dordrecht: Kluwer), p. 515
\ref Kurfess, J.D. et al. 1992, ApJ, 399, L137
\ref Li, W.D. et al. 1999, in Proceedings of the 10th 
Maryland Astrophysics Conference, ed. S.S. Holt \& W.H.Zhang, 
(New York:AIP), p.91
\ref Lin, H. et al. 1996, ApJ, 464, 60
\ref Livio, M. 2000, astro-ph, 0005344, 22 pages
\ref Marzke, R.O., et al. 1998, ApJ, 503, 617
\ref Matz, S.M. et al. 1988, Nature, 331, 416
\ref Milne, P. A., The, L.-S., and Leising, M. D. 1999, ApJS, 124, 503
\ref Morris, D.J. et al. 1997, in Proceedings of the 4th Compton Symposium, 
ed. C.D. Dermer, M.S. Strickman, J.D. Kurfess, (New York:AIP), p.1084
\ref Nomoto, Thielemann, F.-K., Yokoi, K. 1984, ApJ, 286, 644
\ref Nugent, P. et al. 1995, ApJ, 441, L33
\ref Oberlack, U. et al. 1996, A\&AS, 120C, 311
\ref Pinto, P.A., Woosley, S.E. 1988, ApJ, 329, 820
\ref Pinto, P.A., Eastman, R.G. 2000a, ApJ, 530, 744 
\ref Pinto, P.A., Eastman, R.G. 2000b
\ref Pinto, P.A., Eastman, R.G., Rogers, T. 2000, 
\ref Pozdnyakov, L. A., Sobol, I. M., and Sunyaev, R. A. 1983,
Ap. Space. Phys. Rev., 2, 189
\ref Ruiz-Lapuente, P., and Spruit, H. 1998, ApJ, 500, 360
\ref Sch\H{o}nfelder, V. et al. 1999, in Proceedings of the 5th Compton Symposium,
ed. M.L. McConnell \& J.M. Ryan, (New York:AIP), p.54
\ref The, L.-S., Bridgman, W. T., and Clayton, D. D. 1994, ApJS, 93, 531
\ref The, L.-S., Burrows, A., and Bussard, R. W. 1990, ApJ, 352, 731
\ref The, L.-S., Clayton, D. D., and Burrows, A. 1990, in IAU Symp. 143,
Wolf-Rayet Stars in Galaxies, ed., K. A. van der Hucht \& B. Hidayat
(Dordrecht: Kluwer), 537
\ref Truran, J.W., Arnett, W.D., Cameron, A.G.W. 1967, Can. J. Phys., 
45, 2315
\ref Wheeler, J.C. 1995, in Proceedings of the NATO ASI on 
Evolutionary Processes in Binary Stars (C477), 
ed. R.A.M.J. Wijers, M.B. Davies, C.A. Tout, (Dordrecht: Kluwer), p. 307
\ref Woosley, S. E.,  Pinto, P. A., and Hartmann, D. H. 1989, ApJ, 346, 395
\ref Yamaoka, H., et al. 1992, ApJ, 393, L55
}
\end{document}